\documentclass[12pt]{JHEP3}

\usepackage{epsfig}

\def \as {\relax\ifmmode\alpha_s\else{$\alpha_s${ }}\fi}
\def \LQCD {\Lambda_{\mbox{\tiny QCD}}}

\def\ba {\begin{eqnarray}}
\def\ea {\end{eqnarray}}

\def\be {\begin{equation}}
\def\ee {\end{equation}}
\def\stot {{1\over \sigma_{\rm tot}}}

\def \pt {{\rm PT}}
\def \np {{\rm NP}}

\title{Scaling Rule for
Nonperturbative Radiation  
in a Class of Event Shapes}

\author{Carola F. Berger\\ C.N.\ Yang Institute for Theoretical Physics,
Stony Brook University, SUNY Stony Brook, NY 11794-3840, U.S.A.\\ and \\
I.N.F.N. Sezione di Torino, Via P. Giuria 1, I-10125 Torino, Italy\thanks{Address after Sept. 1st, 2003.}\\
E-mail: \email{carola.berger@to.infn.it}}
\author{George Sterman\\C.N.\ Yang Institute for Theoretical Physics,
Stony Brook University, SUNY Stony Brook, NY 11794-3840, U.S.A.\\ 
E-mail: \email{sterman@insti.physics.sunysb.edu}}

\abstract{We discuss nonperturbative radiation for
a recently introduced class of infrared safe event shape weights,
which describe the narrow-jet limit.
Starting from next-to-leading logarithmic (NLL) resummation,
we derive an approximate scaling rule that relates the
nonperturbative shape functions for these weights 
to the shape function for the thrust.
We argue that the scaling reflects the boost invariance
implicit in NLL resummation, and discuss its limitations.
In the absence of data analysis 
for the new event shapes, we compare these predictions to the output of
the event generator PYTHIA.}


\preprint{YITP-03-36\\ \today}

\begin{document}

\section{Introduction}
\label{sec:intro}

The hadronic final states of hard-scattering reactions
encode information about the full range of QCD dynamics,
from the short-distance production of quarks and gluons
reflected in jet production,
to the long-time process of hadronization reflected
in the exclusive distributions of particles in the final state.  Event
shapes  \cite{Farhi:1977sg,weights,broad1,heavyjet}, which generalize jet cross sections, are
functions of final-state momenta
that measure the flow
of energy \cite{eflow}.  We refer to these functions as
`weights' below. Event shape weights are
tools with  which we may analyze the long distance behavior
of QCD through jet events.
Qualitatively, we expect that cross sections for states with narrow, low-mass jets  
are sensitive to long-distance dynamics, while more inclusive 
cross sections, dominated by higher-mass jets, are
predicted more accurately by
perturbation theory.  By varying the values of event shape weights
we may select narrow or wide jet events, and thus vary the relative importance of
long- and short-distance dynamics.  

Infrared safe event shapes are those
that are well-defined order by order in perturbation theory.   
For the mean values of such event shapes, or for
values of the weight where higher order perturbative corrections
are relatively small, nonperturbative
effects enter as additive power corrections.
The analysis of perturbation theory suggests that 
these  are
typically integer powers of $1/Q$, with $Q$ the overall 
center of mass (c.m.) energy
in annihilation processes or the momentum transfer
in deep-inelastic scattering 
\cite{irren,Manohar:1994kq,irrdiff,DokWeb,dispers,oneqdisrefs,oneqdish1}.  
The universality properties
of power corrections, also inferred from perturbation theory,
can be used to provide measurements of the strong
coupling \cite{alphasmean,MovillaFernandez:2001ed}.  As we approach the narrow jet
limit, however, perturbative cross sections generically develop
large logarithmic corrections.

When the weights of event shapes
are symmetric in the phase space of final state particles, 
their large logarithmic corrections may be resummed 
in the limit of narrow jets, to next-to-leading
logarithm (NLL)  \cite{CTTW1,CTTW2,Dokshitzer:1998kz}
and beyond \cite{BKuS} in $\rm e^+e^-$ annihilation.  
The analogous but even  more challenging analysis has been
carried out for jet events in deep-inelastic scattering \cite{dsdis,dsnonglo}.
For many such quantities in $\rm e^+e^-$ annihilation, it is possible to
factorize perturbative and nonperturbative corrections in a well-defined
fashion \cite{KorSt99}, and to study the latter phenomenologically.

Familiar examples include the thrust, $T$ \cite{Farhi:1977sg}, jet
broadening, $B$ \cite{broad1}, and the heavy jet mass, $\rho_H$ \cite{heavyjet}.  Short
distance effects are treated in resummed perturbation
theory and control the leading-power
behavior in $Q$
including logarithms \cite{CTTW1,CTTW2,Dokshitzer:1998kz,BKuS}.
Long-distance effects still enter as power corrections,
whose form may be inferred  
from perturbation theory in a manner analogous
to power corrections to the mean values of event shapes \cite{irrdiff,DokWeb}.  
  An example is $1-T$, with $T$ the thrust in $\rm e^+e^-$ annihilation, for which
the first power correction near $T=1$ is $1/[(1-T)Q]$.
This correction grows rapidly as $T\rightarrow 1$, but it
exponentiates  \cite{irrdiff,DokWeb}.  The effect is to
shift the perturbative distribution \cite{irrdiff,DokWeb}, 
which allows 
additional determinations of the strong coupling 
in conjunction with nonperturbative parameters \cite{MovillaFernandez:2001ed}.
 For small $1-T$, not
only the first power correction, but 
all corrections of the form $1/[(1-T)Q]^n$ may
be organized into shape functions \cite{KorSt99,KorMor98} 
in a manner that we will review below.  

Considerable progress has been made in
determining shape functions for the thrust, C-parameter and heavy jet mass
\cite{Korchemsky:2000kp,BKoS,GarRat1,GarRat2,GarMag}.
It is clearly desirable, however, to widen the class of
event shapes to allow systematic studies of 
the approach to nonperturbative dynamics in
infrared safe observables.  

In this paper, we
attempt to pursue such an approach, using the
class of event shapes introduced in \cite{BKuS},
which includes the thrust and jet broadening as
special cases.  
As in \cite{BKuS}, we will restrict ourselves to
event shapes that vanish in the two-jet limit for
$\rm e^+e^-$ annihilation, but extensions
to quantities that probe multi-jet
structure are possible \cite{BSZ},
and we will discuss the differences 
briefly below.

We will follow Refs.\ \cite{irrdiff,DokWeb,KorSt99}
in deducing the form of power corrections from
the resummed event shape weights at next-to-leading
logarithm \cite{BKuS} in terms of shape functions.
For the class of weights
that we consider, we will identify  an approximate scaling
relation for their nonperturbative shape functions.  
We argue that this scaling reflects the boost
invariance of the underlying soft dynamics.
We will work for the most part directly in a
transform space, where the resummation of large
perturbative as well as power corrections is
most transparent.

Because we know of no experimental data analysis for
this class of observables,
we compare the predictions
of our scaling rule to the event generator PYTHIA \cite{Sjostrand:2000wi}.
As we shall see, PYTHIA reproduces the scaling
over a range of parameters.  We will briefly discuss
the dynamical effects that are neglected in the
derivation of our scaling law, and what
we may conclude from the event generator results.

The following section reintroduces the event shapes of Ref.\ \cite{BKuS},
and in Section 3 we review their resummation, including their
numerical evaluation in perturbation theory, following
the methods of Refs.\ \cite{CTTW1,CTTW2}.  We derive the
scaling rule and describe its interpretation
in terms of boost invariance in Section 4. In Section 5 we discuss
 its comparison with
PYTHIA \cite{Sjostrand:2000wi} for shifts in the distribution peaks and
for moments of the shape functions.  We close with a 
summary.  A brief appendix gives some details on
the evaluation of integrals for the  resummed
distributions.

\section{The Class of Event Shapes}\label{sec:def}

We consider final states $N$ characterized
by two nearly back-to-back jets,
\be
e^+ + e^- \rightarrow J_1(N) + J_2(N)\, ,
\label{crossdef1}
\ee
at
center of mass  energy $Q \gg \LQCD$. For each
such final state, a thrust axis is determined
as the unit vector $\hat{n}$ that maximizes the
the thrust $T$ \cite{Farhi:1977sg}:
\be
T(N) = \max\limits_{\hat{n}} \frac{\sum\limits_{{\rm all}\ i\in N} 
\left| \vec{p}_i \cdot
\hat{n}\right| }{\sum\limits_{{\rm all}\ j\in N} \left| \vec{p}_j 
\right|}\, . \label{thrust}
\ee
The class of event shapes $\tau_a$ of Ref.\ \cite{BKuS}
is defined in terms of this axis.
Parameter $a$ is
adjustable, $- \infty < a < 2$, and allows us to study
various event
shapes within the same formalism.  It helps to control the
approach to the two-jet limit. 

The weight functions defined
in Ref.\ \cite{BKuS} for a state $N$ are 
\be
\tau_a(N) = {1\over Q}\sum_{{\rm all}\ i\in N}\ \omega_i\, \left(\sin \theta_i\right)^a\, 
\left( 1 - |\cos \theta_i|\right)^{1-a}\, ,
\label{weightzero}
\ee
in terms of particle energies $\omega_i$ and the angles $\theta_i$ 
of the particle momenta to the thrust axis.  For
the discussion below, we will find it more convenient to
introduce a form directly in terms of particle momenta,
\be
\tau_a(N)
= {1\over Q}\sum_{{\rm all}\ i\in N}\ p_{i\; T}\; e^{-|\eta_i|(1-a)}\, ,
\label{barfdef}
\ee
where $p_{i\; T}$ is the transverse momentum relative to the 
thrust axis, and $\eta_i$ is the corresponding pseudorapidity,
\be
\eta_i = \ln \cot \left(\theta_i/2\right)\, . \label{rapidity}
\ee
The case $a=0$ in Eq.\ (\ref{barfdef}) is essentially $1-T$, with $T$
the thrust while $a=1$ is the jet  broadening.
A similar weight function with a non-integer power has been discussed in
a related context for $2>a>1$ in
\cite{Manohar:1994kq}.
The two forms for the weight functions, Eqs.\ (\ref{weightzero}) and (\ref{barfdef}) are 
equivalent for massless particles.
We will use (\ref{barfdef}), in terms of momenta and pseudorapidities, in the
analysis of nonperturbative corrections below.
We will recall later
the difference between the two forms of the event shapes when applied 
to massive hadrons \cite{Salam:2001bd}.

Any choice $a<2$ in (\ref{barfdef}) specifies an infrared safe event shape
variable, because the
contribution of any particle $i$ to the event shape behaves as
$\theta_i{}^{2-a}$ in the collinear limit, $\theta_i=\cos^{-1} |\hat
n_i\cdot \hat n|
\rightarrow 0$. However, the resummed formula given below from Ref.\
\cite{BKuS} is
valid only for $a < 1$. For
$a \geq 1$ recoil effects have to be taken into account, at
least beyond the level of leading logarithm \cite{Dokshitzer:1998kz}.
As $a \rightarrow 2$ the weight vanishes only  very slowly for
$\theta_i \rightarrow 0$, and at fixed $\tau_a$, the
jets become very narrow. On the other hand, the limit $a\rightarrow -\infty$
corresponds to the total cross section.

The  differential cross section
for such dijet events at fixed values of $\tau_a$ is given by
\ba
{d \sigma(\tau_a, Q)\over
d\tau_a}
&=&
{1\over 2 Q^2}\ \sum_N\;
|M(N)|^2\,  \delta(\tau_a - \tau_a(N)),
\label{eventdef}
\ea
where we sum over all final states $N$ that contribute to the
weighted event, and where $M(N)$ denotes
the corresponding amplitude for ${\rm e^+e^-}\rightarrow N$.

    Since we are investigating  two-jet cross sections, we fix the
constant $\tau_a$ to be
    much less than unity:
\be
0 < \tau_a \ll 1\, .
\label{elasticlim}
\ee
For small $\tau_a$, the cross section
(\ref{eventdef}) has
corrections in
$\ln (1/\tau_a)$, which have been organized in Ref. \cite{BKuS}.
In the following we will quote the result of the resummation of large
logarithms of $\tau_a$ in Laplace moment space. The Laplace transform
of the cross section (\ref{eventdef}) is given by
\ba
      \tilde{\sigma} \left(\nu,Q,a \right) & = &  \int^1_0 d \tau_a\, {\rm e}^{\;
-\nu\,
\tau_a}\ {d
\sigma(\tau_a,Q) \over d \tau_a}\, .
\label{trafo}
\ea
Logarithms of $1/\tau_a$ are transformed to logarithms of $\nu$.
For large $\nu$, dependence on the upper limit in the $\tau_a$ 
integral is exponentially suppressed.
Here and below quantities with tildes are the transforms in
$\tau_a$,
and quantities without tildes denote untransformed functions.
Our results  below are valid in the region where $\ln \nu$ is much
larger than $|a|$ \cite{BKuS}.

\section{The Resummed Cross Section at NLL}

\subsection{The resummed cross section in moment space}

\begin{samepage}
The NLL resummed cross section (\ref{trafo}) for $a < 1$ in
moment space can be written as \cite{BKuS}
\ba
\hspace*{-5mm} {1\over \sigma_{\rm tot}} \, \tilde{\sigma}
\left(\nu,Q,a \right) \!
&=& \!
      \exp \Bigg\{ 2\, \int\limits_0^1 \frac{d u}{u} \Bigg[ \,
      \int\limits_{u^2 Q^2}^{u Q^2} \frac{d p_T^2}{p_T^2}
A\left(\as(p_T)\right)
      \left( {\rm e}^{- u^{1-a} \nu \left(p_T/Q\right)^{a} }-1 \right)
\nonumber \\
      & & \qquad \qquad \quad \quad
      + \frac{1}{2} B\left(\as(\sqrt{u} Q)\right) \left( {\rm e}^{-u
\left(\nu/2\right)^{2/(2-a)} } -1 \right)
      \Bigg] \Bigg\} 
\nonumber\\
 & \equiv &
\left[\,{\mathcal{J}}(\nu,Q,a)\right]^2\, ,
\label{thrustcomp}
\ea
where ${{\mathcal{J}}}(\nu,Q,a)$ 
is a factorized function associated with each jet. 
The resummation is in terms of anomalous dimensions $A(\alpha_s)$ and $B(\alpha_s)$,
which have finite expansions in the  running coupling, 
\ba
A(\as) = \sum_{n=1}^\infty A^{(n)}\ \left({\as\over
\pi}\right)^n\, ,
\ea
and similarly for $B$.
To NLL they are specified by the well-known coefficients,
\ba
A^{(1)}  & = & C_F
, \\
A^{(2)}  & = & \frac{1}{2} C_F \left[ C_A
\left( \frac{67}{18} - \frac{\pi^2}{6} \right) - \frac{10}{9} T_F N_f
\right], \\
B^{(1)} & = & - \frac{3}{2} \, C_F,
\ea
independent of $a$. $C_F$ and $C_A$ are the Casimir charges of the
fundamental and adjoint representation of SU($N_c$), respectively,
$N_f$ denotes the number of flavors, and $T_F = 1/2$ is the usual
normalization of the generators of the fundamental representation.
Eq.\ (\ref{thrustcomp}) reproduces the
NLL resummed thrust cross section \cite{CTTW1,CTTW2} when $a = 0$.\end{samepage}

\subsection{Inversion of the transform} \label{sec:inv}

As it stands, the resummed cross section (\ref{thrustcomp})
is ambiguous, because of the singularity of the perturbative
running coupling.
To define the integrals in
(\ref{thrustcomp}) and to invert the transformed cross section from moment space
back to $\tau_a$, 
 we will follow the method of Ref.\ \cite{CTTW2}.
In this approach, we avoid the singularities of the
perturbative running coupling by reexpressing the
running coupling in terms of the coupling at a hard scale,
and by performing the resulting integrals in the exponents
of Eq.\ (\ref{thrustcomp}) to NLL.  Explicit expressions for
the cross section, and hence the jet functions $\mathcal{J}$,
are given in Appendix \ref{app:explicit}.  The inversion is then also
carried out to NLL.  In the
final expressions, the singularities of the running coupling
are only manifested as singularities at small values of $\tau_a$.

As in Ref.\ \cite{CTTW2}, we work with
the integrated cross section, also called the radiator,
\ba
    R(\tau_a,Q)  \equiv  \stot \int\limits_0^{\tau_a} d \tau_a'\, {d
\sigma(\tau_a',Q)\over
d\tau_a'}\, .
\ea
The radiator can be found 
directly from the jet functions, $\mathcal{J}(\nu,Q,a)$  in transform space
by
\ba
 R(\tau_a,Q)
&=&    \frac{1}{2 \pi i} \,\int_{C} {d\nu} \, {\rm e}^{\nu  \tau_a}
\; \tilde R(\nu,Q,a)
\nonumber\\
&=& \frac{1}{2 \pi i} \,\int_{C} \frac{d\nu}{\nu} {\rm e}^{\nu
\tau_a} \left[\,{\mathcal{J}}(\nu,Q,a)\right]^2
 \, .
\label{intinv}
\ea
The contour $C$ lies in the complex plane to the right of all
singularities of the integrand, and
the first equality
defines the radiator in transform space.
The differential cross section (\ref{eventdef}) is easily obtained
once we have an explicit form for the radiator,
\be
    \stot {d \sigma(\tau_a,Q)\over
    d \tau_a}  =  \frac{1}{\tau_a} \frac{d}{d \ln \tau_a} R(\tau_a,Q)\,  .
\label{crossdiff}
    \ee

To perform the integral in
Eq.\ (\ref{intinv}), we Taylor expand the resummed exponent with respect to
$\ln \nu$ around $\ln \nu = \ln (1/\tau_a)$, because the functions
$g_1$ and $g_2$ in $\mathcal{J}$ (see Appendix \ref{app:explicit})
vary more slowly with $\nu$ than $\nu \tau_a$ \cite{CTTW2}.
At NLL accuracy we can neglect all derivatives higher than first
order in the Taylor series of the exponent. Performing the integral
is then straightforward, using
\be
\frac{1}{2 \pi i} \int_C du  \,{\rm e}^{u- (1-\gamma) \ln u} =
\frac{1}{\Gamma(1-\gamma)}.
\ee
In this way, we find
\ba
R(\tau_a,Q) & = &  \frac{\exp \Bigg\{ 2 \ln \left(\frac{1}{\tau_a}\right) \, g_1
(x,a) + 2 g_2 (x,a) + 2 (2-a) x^2 \ln\left(\frac{2 \mu}{Q} \right)
g'_1 (x,a) \Bigg\} }{\Gamma\Bigg[1-2 g_1 (x,a) - 2 x
g'_1(x,a)\Bigg]}\, . \nonumber \\
& &
\label{crossintsol}
\ea
The functions $g_i,i = 1,2,$ and $g'_1$ are given in Eqs.\
(\ref{g1}), (\ref{g2}), and (\ref{g1prime}) of Appendix \ref{app:explicit}
in terms of the variable
\be
x \equiv \frac{\as(\mu)}{\pi} \frac{\beta_0}{2\,(2-a)} \ln
\left(\frac{1}{\tau_a}\right)\, , 
\label{xdef}
\ee
with $\beta_0$ the first coefficient of the QCD beta function.
The explicit formulas in the appendix show that the $g_i(x,a)$ have
logarithmic singularities at $x=1$ and $x=1/(2-a)$,
which are the manifestations of the 
singularities of the perturbative running coupling
for this NLL evaluation.  Although well defined for most values
of $\tau_a$, the resummed cross section remains ambiguous
beyond NLL.

\subsection{Matching and numerical evaluation}
\label{num}

A realistic evaluation of the resummed cross section requires
matching to fixed-order calculations. While the resummed predictions are
reliable for small values of $\tau_a$, fixed-order contributions are more accurate
at larger $\tau_a$.
For matching to NLL accuracy, it is necessary to know the fixed order
contributions up to ${\mathcal{O}}(\as^2)$.
These can be calculated numerically using, for example, the program
EVENT2 \cite{Catani:1996jh}.

In the following we will use log $R$ matching \cite{CTTW2}.  Other matching schemes are possible, and
differ from this formula at order $\as^3$ and
NNLL. The
radiator at NLL, evaluated at the scale $\mu  = Q$, is computed as
\be
\ln R^{\mbox{\tiny NLL}} \left(\tau_a,Q \right)
= \ln R_{\mbox{\tiny resum}}
\left(\tau_a,Q \right) + 
\ln R_{\mbox{\tiny fixed}}
\left(\tau_a,Q  \right) - \ln R_{\mbox{\tiny
resum}}^{\mbox{\tiny exp}} \left(\tau_a,Q
\right)   \, . \label{match}
\ee
Here the first term on
the right, $\ln R_{\mbox{\tiny resum}}$, is the logarithm of the resummed radiator, 
which can be read off of Eq.\ (\ref{crossintsol}). 
$\ln R_{\mbox{\tiny fixed}}$ is the logarithm of the fixed order radiator calculated with
EVENT2 at NLO, expanded to order $\as^2$,
\be
\ln R_{\mbox{\tiny fixed}}
\left(\tau_a,Q  \right) \equiv \left(\frac{\as(Q)}{\pi}\right) R_{\mbox{\tiny fixed}}^{(1)} +
\left(\frac{\as(Q)}{\pi}\right)^2 \left[ R_{\mbox{\tiny fixed}}^{(2)} - \frac{1}{2} 
\left( R_{\mbox{\tiny fixed}}^{(1)}\right)^2 \right],
\ee
where the $R_{\mbox{\tiny fixed}}^{(i)},\,i = 1,2$ are the first- and second-order parts, 
respectively,
of the radiator given by EVENT2.
The last term on the right of Eq. (\ref{match}), 
 $\ln R_{\mbox{\tiny
resum}}^{\mbox{\tiny exp}} \left(\tau_a,Q \right) $, 
is the logarithm of the resummed radiator
expanded to order $\as^2$, which needs to be subtracted in order to avoid double counting. This
contribution is found from Eq. (\ref{crossintsol}) with Eqs.\
(\ref{g1}), (\ref{g2}), and (\ref{g1prime}) of Appendix \ref{app:explicit}:
\ba
\ln R_{\mbox{\tiny
resum}}^{\mbox{\tiny exp}} \left(\tau_a,Q
\right) & \equiv & \left(\frac{\as(Q)}{\pi}\right) \left[ G_{11}(a) \ln \left( \frac{1}{\tau_a} \right) + 
G_{12}(a) \ln^2 \left( \frac{1}{\tau_a} \right) \right] \nonumber \\
& + & \left(\frac{\as(Q)}{\pi}\right)^2 \left[ G_{22}(a) \ln^2 \left( \frac{1}{\tau_a} \right) + 
G_{23}(a) \ln^3 \left( \frac{1}{\tau_a} \right) \right], \label{resumexp}
\ea
the functions $G_{ij}$ are listed in the appendix. Finally, the physical requirement
that the cross section vanishes beyond the upper kinematic boundary, $\tau_a^{\mbox{\tiny max}}$, 
imposes the following constraints on the matched expression (\ref{match}),
\be
R^{\mbox{\tiny NLL}}\left(\tau_a^{\mbox{\tiny max}},Q\right) = 1, \qquad  \left.\frac{d R^{\mbox{\tiny NLL}}\left(\tau_a,Q\right)}{d \tau_a} \right|_{\tau_a = \tau_a^{\mbox{\tiny max}}} = 0. \label{constraint}
\ee
This is achieved by the replacement \cite{CTTW2}
\be
\frac{1}{\tau_a} \rightarrow \frac{1}{\tau_a} - \frac{1}{\tau_a^{\mbox{\tiny max}}} + 1 \label{maxreplace}
\ee
in the resummed terms, since the fixed order coefficients satisfy Eqs. (\ref{constraint}) order by order. 
$\tau_a^{\mbox{\tiny max}}$ depends on the number of final-state particles, corresponding to the order
in $\as$ up to which the fixed order contributions are evaluated. The replacement (\ref{maxreplace})
suppresses terms at large $\tau_a$ higher than this order that are generated 
inaccurately by the resummed contribution. 
$\tau_a^{\mbox{\tiny max}}$ at the leading logarithmic level is given by the limiting configuration
of three final-state particles  \cite{Berger:2003zh},
\be
\tau_a^{\mbox{\tiny max},LL} = \frac{\sqrt{3}^a}{3}.
\ee
The limit from four particles in the final state 
which gives the upper kinematic boundary at NLL can easily be determined from the output of EVENT2. 

Eq.\ (\ref{crossintsol}) for the radiator and
hence  (\ref{crossdiff}) for the cross sections, is 
applicable, with power and subleading logarithmic corrections
for values of $\tau_a$ away from the end-point region,  where 
the variable $x$ in Eq.\ (\ref{xdef})
becomes of order unity.  That is, we require
$[\beta_0/(2 \pi)] \as \ln (1/\tau_a)$ $< 1$, or equivalently $\tau_a > \LQCD/Q$.
   For $\tau_a \sim \LQCD/Q$
non-perturbative corrections become dominant. 
In this range of $\tau_a$, we ``freeze''
the perturbative contribution to the cross section,
following \cite{KorMor98,Korchemsky:2000kp},
\be
R_{\pt} \left(\tau_a,Q,\kappa \right) \equiv \theta\left(\tau_a -
\frac{\kappa}{Q} {\rm e}^{-\gamma_E} \right) R^{\mbox{\tiny NLL}}
\left(\tau_a,Q \right) + \theta\left(\frac{\kappa}{Q}
{\rm e}^{-\gamma_E} -\tau_a  \right) R^{\mbox{\tiny NLL}}
\left(\frac{\kappa}{Q},Q \right),
\label{freeze}
\ee
where $R^{\mbox{\tiny NLL}}$ is evaluated according to (\ref{intinv})
and (\ref{match}), and where $\kappa$ is a nonperturbative cutoff. Dynamics below
the scale $\kappa$ will be incorporated into the nonperturbative corrections, 
in a manner we 
will discuss below. For our numerical
studies we pick $\kappa = 0.75$ GeV, as in \cite{KorMor98}.

\section{The Scaling Rule}

We are now ready to derive the scaling relation for
nonperturbative shape functions.  We show first how
the rule is implied by the resummed cross section that
we have just described, and go on to interpret the 
physical content of the scaling.

\subsection{From resummations to shape functions}

Following Ref.\ \cite{KorSt99}, we identify the
power structure of nonperturbative corrections by
a direct expansion of the integrand
in the resummed exponent
at momentum
scales below an infrared factorization scale, $\kappa$.
Although this scale need not be exactly the same
as the scale in Eq.\ (\ref{freeze}) at which the radiator is frozen,
they are closely related, and we will use the same symbol for both.
We thus rewrite Eq.\ (\ref{thrustcomp})  as
the sum of a perturbative term, summarizing all $p_T>\kappa$,
and a soft term, containing the nonperturbative
physics of strong coupling.  This
corresponds to $p_T<\kappa$ in the term with anomalous dimension $A$,
and $uQ^2<\kappa^2$ in the term with $B$.  Exchanging the
order of integration for the $A$ term, we find
\ba
\vspace*{-1mm} \ln  \tilde R(\nu,Q,a)
&=&    \ln  \tilde R_{\pt}(\nu,Q,\kappa,a)
\nonumber\\
     &\ &
+      2\, \,
      \int\limits_{0}^{\kappa^2} \frac{d p_T^2}{p_T^2}
A \left(\as(p_T)\right)\ \int\limits_{p_T^2/Q^2}^{p_T/Q} \frac{d u}{u}
      \left( {\rm e}^{- u^{1-a} \nu \left(p_T/Q\right)^{a} }-1 \right)
\nonumber \\
       & & \qquad \qquad \quad \quad
+  \int\limits_0^{\kappa^2/Q^2} {du\over u} B\left(\as(\sqrt{u}
Q)\right) \left( {\rm e}^{-u
\left(\nu/2\right)^{2/(2-a)} } -1 \right)
\nonumber\\
     &=&   \ln  \tilde R_{\pt}(\nu,Q,\kappa,a)
\nonumber\\
& &  +\,
\frac{2}{1-a}\
\sum_{n=1}^\infty\ \frac{1}{n\, n!}\, {\left(-{\nu\over Q}\right)}^{n}
\! \int\limits_{0}^{\kappa^2} {dp_T^2\over p_T^2}\; p_T^n\;
A\left(\alpha_s(p_T)\right) \!
\left[ 1 - \left({p_T \over Q}\right)^{n(1-a)}\right]
\nonumber\\
     &\ &  +\,  \sum_{n=1}^\infty (-1)^n 
\frac{\left(\nu/2\right)^{2n/(2-a)} }{n!}\,
\int\limits_0^{\kappa^2/Q^2} \frac{du}{u}\, B\left(\as(\sqrt{u} Q)\right)\; 
u^n
\nonumber\\
&\  \equiv &
\ln \tilde R_{\pt}(\nu,Q,\kappa,a)
+ \ln \tilde f_{a,\np}\left(\frac{\nu}{Q},\kappa \right)
 \nonumber \\
 & & \qquad \qquad \qquad \qquad \qquad \qquad + {\cal O}\left( \nu\left({\kappa\over 
Q}\right)^{2-a},\nu^{\frac{2}{2-a}}\left({\kappa\over 
Q}\right)^{2}\right)\, .
\label{pcshape}
\ea
In the second equality we have expanded the exponentials and integrated over $u$. In the
last equality we have introduced the logarithm
of the shape function, as an expansion in powers of $\nu/Q$,
\ba
\ln \tilde f_{a,\np}\left(\frac{\nu}{Q},\kappa \right)
&=&
\,
\frac{2}{1-a}\
\sum_{n=1}^\infty\ \frac{1}{n\, n!}\, {\left(-{\nu\over Q}\right)}^{n}
\int\limits_{0}^{\kappa^2} {dp_T^2\over p_T^2}\; p_T^n\;
A\left(\alpha_s(p_T)\right)\
\nonumber\\
&\equiv& \frac{1}{1-a}\; \sum_{n=1}^\infty\ \lambda_n(\kappa)\,
{\left(-{\nu\over Q}\right)}^{n}
\, .
\label{fdef}
\ea
We neglect the terms $(p_T/Q)^{n(1-a)}$
in Eq.\ (\ref{pcshape}) for the 
expansion with $A$, and the entire $B$ expansion, as
indicated. These terms are suppressed by additional (fractional) powers of $Q$
only,
of course, for $a<1$.  This is the same
restriction to the NLL resummation formula, Eq.\
(\ref{thrustcomp}) that follows from considerations of recoil \cite{BKuS}.
Given this approximation, we find the simple result
that the only dependence on $a$ is through an overall factor $1/(1-a)$.

Equation (\ref{fdef}) immediately leads to the following scaling relation between
shape functions for different values of $a$:
\be
\ln\, \tilde f_{a,\np}\left(\frac{\nu}{Q},\kappa \right) = {1-b\over 
1-a}\; \ln\, \tilde
f_{b,\np}\left(\frac{\nu}{Q},\kappa \right)\, .
\label{freln}
\ee
We can now use the shape
function determined by the thrust ($a=0$) to predict shape
functions for any value of $a<1$.  Of course, for $a>0$,
the relative suppression of the neglected
terms in Eq.\ (\ref{pcshape})
is by fractional powers, and we will apply our results
only to $a\le 0$ below.  It is worth noting that
for $2>a>1$, where the event shape weight
vanishes very slowly in the collinear limit, the
fractional powers in (\ref{pcshape}) dominate the
integer powers, as observed in Ref.\ \cite{Manohar:1994kq}.
This, however,  is outside
the range of $a$ to which our formalism applies at NLL \cite{BKuS}.

\subsection{Interpretation of the scaling}
\label{origin}

The assumptions that entered our result (\ref{freln}) are
relatively standard: that nonperturbative parameters
may be identified with moments of the running coupling
with respect to its scale,
or more generally moments of the anomalous dimension $A(\alpha_s)$.  Our
analysis does not predict the values of the nonperturbative parameters
$\lambda_n$ aside from their overall $a$-dependence,
and is consistent with the absence of even power corrections, as argued in
Refs.\ \cite{GarRat1,GarRat2,GarMag}.  It is also based on the specific
form of NLL resummation.
Despite these limitations, we can identify the
physical origin of the scaling.  We have seen 
that the leading powers come entirely from the $A(\alpha_s)$
term in the exponent of the resummed NLL cross section.  This 
contribution
can be derived directly from the eikonal approximation,
in  which all soft gluons are emitted from light-like
 Wilson lines.

To be specific, we can define the eikonal cross section by
\begin{equation}
\label{NonOPEf}
\frac{d \sigma^{\rm (eik)}(\tau_a,Q)}{d \tau_a}
= \sum_{N} |\langle N | U (0) | 0 \rangle |^2
\delta \left( \tau_a-\tau_a(N) \right)\,  \theta\left(Q-E(N)\right)\, ,
\end{equation}
where $\tau_a(N)$ is the weight computed
according to Eq.\ (\ref{barfdef}) for state $N$, and $E(N)$ is the
total energy of the  particles in $N$.  The operator $U(0)$
is defined in terms of  Wilson lines,
${\mit\Phi}_{\beta}^{(f)} [ \infty, y ] = P \exp \left[ i g
\int_{0}^{\infty} d \sigma \, \beta^\mu A_\mu^{(f)} (\sigma \, \beta + y) \right]$,
where $(f)$ labels the representation (quark or antiquark in our case)
and where $P$ indicates color ordering along the corresponding
light-like direction $\beta$.  In this notation, we define
\begin{eqnarray}
\label{Pexponent}
U (0)
= T\, \left[\, {\mit\Phi}_{\beta_1}^{(q)} [ \infty, 0 ]\;
{\mit\Phi}_{\beta_2}^{(\bar q)} [ \infty, 0 ]\, \right]
\, .
\end{eqnarray}
with $\beta_1^2=\beta_2^2=0$ and $\beta_1\cdot \beta_2=1$,
and with $T$ the time ordering operator.
We work in a frame where $\vec \beta_1$ and $\vec \beta_2$
are back-to-back.  

Diagrammatic rules for constructing the perturbative expansion
of Eq.\ (\ref{NonOPEf}) may be found, for example, in Ref.\ 
\cite{eikonalrules}, and
the exponentiation properties 
applicable to logarithmic corrections for eikonal event shapes
 in Refs.\ \cite{eikexp,Berger:2003zh}.
Here we only need to stress
that matrix elements computed from the product of back-to-back
Wilson lines in Eq.\ (\ref{NonOPEf}) are boost invariant\
along  the axis defined by $\vec \beta_1$ and $\vec \beta_2$ and 
are also invariant under the scaling of
all final state momenta ($p_i\rightarrow \lambda p_i$)  up to
renormalization effects.  Frame dependence
enters the eikonal event shape distributions 
only through fixing the shapes themselves, which
distinguish between positive and negative rapidities (see Eq.\ 
(\ref{barfdef})), and through ratios
of the cutoff in the overall energy, $Q$, to the renormalization scale. 
In particular, when the decay products of a particle are emitted into
both hemispheres \cite{BKoS,nasonseymour}, 
the event shape receives a noninvariant contribution,
even if the amplitude to produce the virtual particle is boost invariant.

These features of the eikonal cross section are illustrated by
the exponent of the resummed perturbative cross section in Eq.\ 
(\ref{thrustcomp}), which reflects
single-particle kinematics in the exponent.
Comparing the definition of the weight
in Eq.\ (\ref{barfdef}) with the NLL resummation (\ref{thrustcomp}),
we can change variables from $u$ to pseudorapidity (\ref{rapidity}), 
by $u=(p_T/Q)\exp(-|\eta|)$,
$du/u=-(|\eta|/\eta)\, d\eta$.
The boost and scale invariance
of the single-particle emission cross section in eikonal approximation
then become manifest.

In general, the interplay between the limits
of integration and the transform links  the
rapidity and transverse momentum integrals in Eq.\  (\ref{thrustcomp}), to
produce the nontrivial $a$-dependence of
the functions $g_i$ in Eq.\ (\ref{crossintsol}), given explicitly in the
Appendix.
When the transverse momentum is limited
by the cutoff $\kappa$ of Eq.\ (\ref{pcshape}),
however,  the dependence on the remaining limits of integration 
is simplified.  To see this, we
expand the exponential of the $A$ term in 
the first equality of (\ref{pcshape}).
For $p_T\le \kappa$, with $\kappa\ll Q$, we have
$1/[n(1-a)]<\ln(Q/p_T)$ for all $n\ge 1$, and the 
$\eta$ integral of the $n$th term in the
expansion is cut off by the exponential $\exp[-n(1-a)|\eta|]$.
Approximating this factor by unity when the exponent is
smaller than one, and by zero elsewhere,
each integral becomes a simple measure of the rapidity
range over which the single-particle weight is negligible,
\ba
\int\limits_{0}^{\kappa^2} \frac{d p_T^2}{p_T^2}
A \left(\as(p_T)\right)\ \int\limits_{0}^{\ln[Q/p_T]} {d \eta}
       \left( {\rm e}^{- \left(\nu p_T/Q\right)\, {\rm e}^{-(1-a)|\eta|}   }-1 
\right)
&\sim&
\nonumber\\
&\ & 
\hspace{-60mm} 
\sum_{n=1}^\infty\ \frac{1}{ n!}\, {\left(-{\nu\over Q}\right)}^{n}
\int\limits_{0}^{\kappa^2} {dp_T^2\over p_T^2}\; p_T^n\;
A\left(\alpha_s(p_T)\right)\, 
\int\limits_{0}^{[n(1-a)]^{-1}} d\eta \, .
\label{boost}
\ea
By comparing the right-hand side here to the 
the right-hand side of the second equality of 
Eq.\ (\ref{pcshape}), we see that
corrections to (\ref{boost}) are precisely the power-suppressed terms
that we have neglected for $a<1$.  The overall decrease with $1/(1-a)$
thus reflects the shrinking rapidity range that is
available in each term due to
the exponential of the weight.   The 
contribution to each power correction from the region of strong
coupling is boost invariant, but the available rapidity range 
decreases uniformly in $(1-a)^{-1}$.
The observed scaling, then, is a reflection of
the underlying boost invariance of the nonperturbative
dynamics.   

Alternatively, we may consider the scaling with $(1-a)^{-1}$
as a test of the use of NLL resummation to determine
shape functions, and of the extent to which boost invariant
dynamics dominates the differential distributions.
We will return to this viewpoint at the end of the following
section, after testing the scaling
rule with the event  generator PYTHIA.

\section{Tests of Scaling}

At present, we know of no data analysis that would
provide experimental tests of our ability to predict
nonperturbative contributions for the class of event shapes
we are considering, aside from the thrust.
In the absence of such an analysis, we will rely on the
event generator PYTHIA \cite{Sjostrand:2000wi}
as a stand-in for experiment.
In this section we will describe two tests of
our scaling rule.  These are, of course, tests
of consistency with PYTHIA, not with nature.
First we discuss our predictions for the shifts \cite{irrdiff,DokWeb} in
the resummed perturbative distributions due to
the first power correction ($\nu/Q$), and then
the full scaling associated with the summation of all powers in
$\nu/Q$.

For convenience, we evaluate the weight functions
  using 
Eq.\ (\ref{barfdef}),
in terms of particle momenta, using the output
of PYTHIA for the thrust axis, which is also 
computed in terms of particle three-momenta.
The derivation of the scaling rule, Eqs.\ 
(\ref{pcshape}) -- (\ref{freln})
does not distinguish between the weight functions expressed in terms 
of energy (\ref{weightzero}) and in
terms of momentum (\ref{barfdef}),
although with massive hadrons in the final state,
the values of $\tau_a$ can be different.
We therefore expect the scaling to hold only
if the weight functions are computed consistently
in terms of energies or momenta.

\subsection{Shifts of the distributions}

Let us first study the shifts of the distributions. Retaining only the
term with $n = 1$ in (\ref{pcshape}), and suppressing the dependence
on $\kappa$, we obtain the cross section in momentum space from Eq.\
(\ref{intinv}),
\be
R(\tau_a,Q) =    \frac{1}{2 \pi i} \,\int_{C} \frac{d\nu}{\nu} 
{\rm e}^{\nu \left(\tau_a - \frac{1}{1-a} \frac{\lambda_1}{Q}\right)}
\left[{\mathcal{J}}_{\mbox{\tiny PT}}(\nu,Q,a)\right]^2 .
    \ee
In this approximation the integrated cross section is shifted 
\cite{irrdiff,DokWeb} to the right from the
perturbatively calculated spectrum by an amount 
\be
\Delta \tau(a,Q) = \frac{1}{1-a} \frac{\lambda_1}{Q}\, . 
\label{shift}
\ee
To the same approximation, this also holds for the differential cross
section (\ref{crossdiff}) for $\tau_a$ not too small.

Here we study the shifts of the peaks, $\Delta \tau_{p}$
\cite{Berger:2003zh}. From Eq.\ (\ref{freln}) we infer that the
shifts of the peaks for different values of $a$ multiplied by $(1-a)$
are the same when measured at the same scale $Q$:
\be
(1-a ) \,\Delta \tau_{p} (a,Q) = (1-b)\, \Delta\tau_{p} (b,Q). \label{scala}
\ee
Let us compare
this prediction, valid at the partonic level, to the corresponding
cross sections
computed by \textsc{PYTHIA} \cite{Sjostrand:2000wi}, version 6.215
\cite{Sjostrand:2001yu}, at the hadronic level, using
\textsc{PYTHIA}'s implementation of the string fragmentation model
\cite{Andersson}. We use the default
settings of the program. The string picture seems to model the
hadronization process successfully, as a comparison with
 recent data for the thrust shows (see e.\,g.
\cite{Berger:2003zh}).  
As noted above, for comparison with \textsc{PYTHIA} we use the
definition of our event shapes in terms of three-momenta,
(\ref{barfdef}), for both the partonic  and the hadronic level. Other
prescriptions are of course possible
\cite{Salam:2001bd}.

We see from Fig.\ \ref{lineshift} that the shifts 
computed this way obey
the scaling quite well in the range between $a=0$ and $a=-1$.
For larger values of $|a|$, the peaks of the cross sections
move into regions where $\tau_a \sim \LQCD/Q$, and where
our analysis is not reliable.

\EPSFIGURE{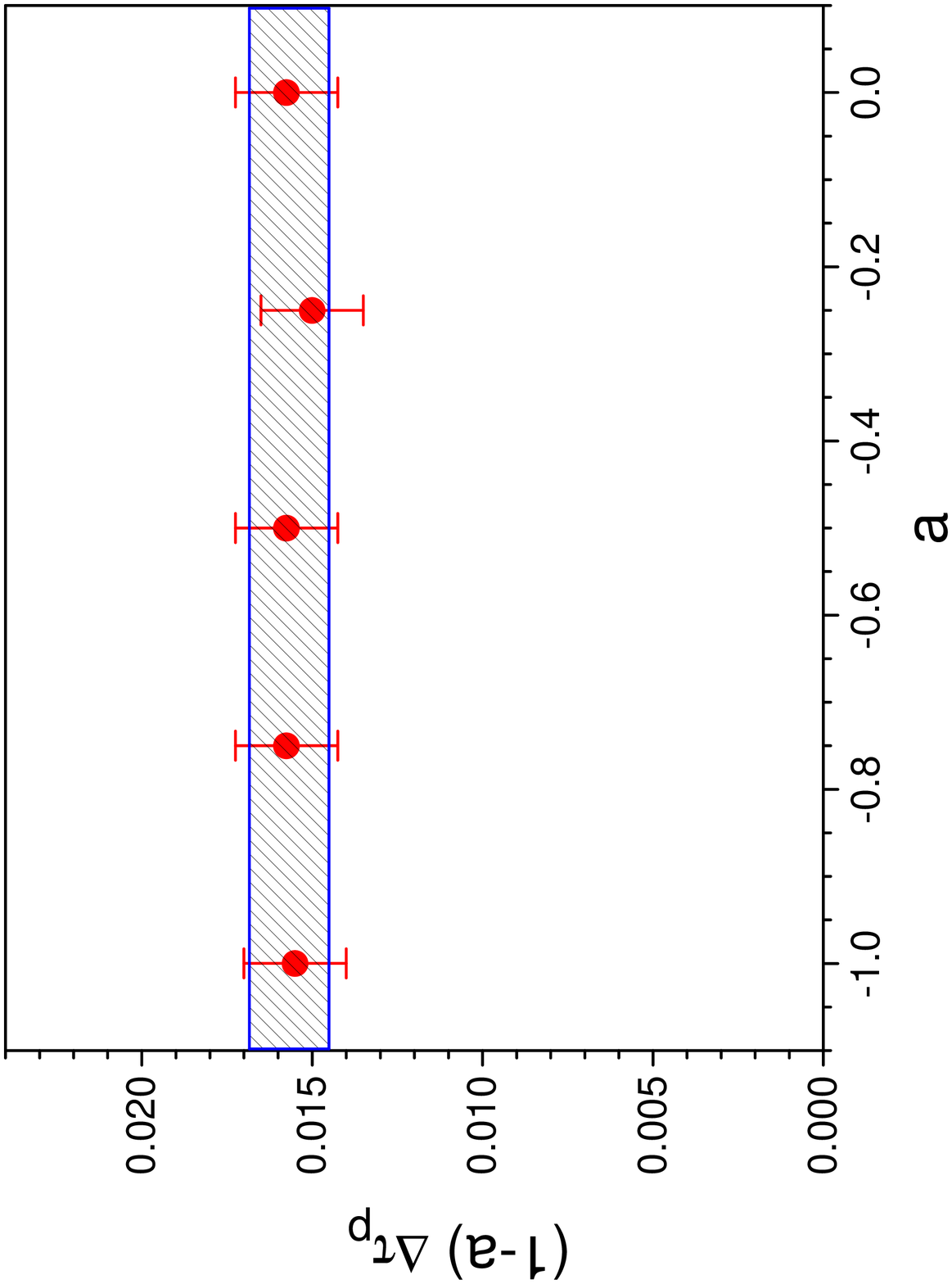,height=10.7cm,angle=270,clip=0}
{Shifts of the peaks $\Delta \tau_{p}(a,Q = 91
\mbox{ GeV})$ of the distributions $\left(1/\sigma_{\mbox{\tiny tot}}\right)
d
\sigma/d \tau_a$ between NLL partonic resummed predictions and
hadronic cross sections computed using \textsc{PYTHIA} with string
fragmentation. The result is multiplied by $(1-a)$. The error bars are estimated
from the uncertainty of the NLL resummed calculation and the output of PYTHIA.
The shaded band is
the shift of the peak for the thrust determined in
\cite{MovillaFernandez:2001ed}  between resummed predictions and
experimental data.  \label{lineshift} }

\subsection{Shape functions}

\begin{samepage}
Now let us turn to the full shape functions.  We will test the scaling
directly in moment
space, where a parameterization in terms of the coefficients
$\lambda_n$ of Eq.\ (\ref{fdef})
is most straightforward.
Using Eq.\ (\ref{pcshape}), we have
\ba
\tilde f_{a,\np}\left(\frac{\nu}{Q},\kappa\right) =
{ \tilde R(\nu,Q,a) \over \tilde
R_{\pt}(\nu,Q,\kappa,a)}
+  {\cal O}\left( {1\over Q^{1-a}}\right)\, .
\label{ratioRs}
\ea\end{samepage}
This gives moment space expressions for the shape functions directly,
given experimental or other input for $\tilde R(\nu,Q,a)$ in
the numerator and resummed perturbation theory (in our case to NLL)
for $\tilde R_{\pt}(\nu,Q,\kappa,a)$ in the denominator.
As above, we use the output of PYTHIA for the numerator.  For the
denominator, we use the method described above in Sec.\ \ref{num}.

The results are shown in Fig.\ \ref{Q91} at $Q=91$ GeV and  in Fig.\ 
\ref{Q35} at $Q=35$ GeV.
We begin by computing the shape functions for $a=0,\, -0.25,\, -0.5$ 
at $Q=91$ GeV and $Q=35$ GeV
directly from Eq.\ (\ref{ratioRs}).  These are the solid curves
in the figures.   The dotted curves show the predictions found by simply
scaling the $a=0$ curve in each case according to Eq.\ (\ref{freln}),
or equivalently
\ba
\tilde f_{a,\np}\left(\frac{\nu}{Q},\kappa\right)
=
\left[\, \tilde f_{0,\np}\left(\frac{\nu}{Q},\kappa\right)\, 
\right]^{1\over 1-a}\, .
\label{rescale}
\ea
The perturbative radiator defined in Eq. (\ref{freeze}),
and thus the ratio in (\ref{rescale}), as plotted in Figs. \ref{Q91} 
and \ref{Q35}, are fairly dependent on the value of the cutoff $\kappa$.
This dependence compensates of course 
between perturbative and nonperturbative 
contributions to the full radiator (\ref{pcshape}).

As in the case of the shifts, we see that within the range of
parameter $a$ considered, the
scaling works well.
  It is worth pointing out that a similar
scaling fails by a relatively large amount for the perturbative cross 
section itself.
We have restricted ourselves to a minimum value of $a=-0.5$ 
to keep an acceptable numerical accuracy.

\EPSFIGURE{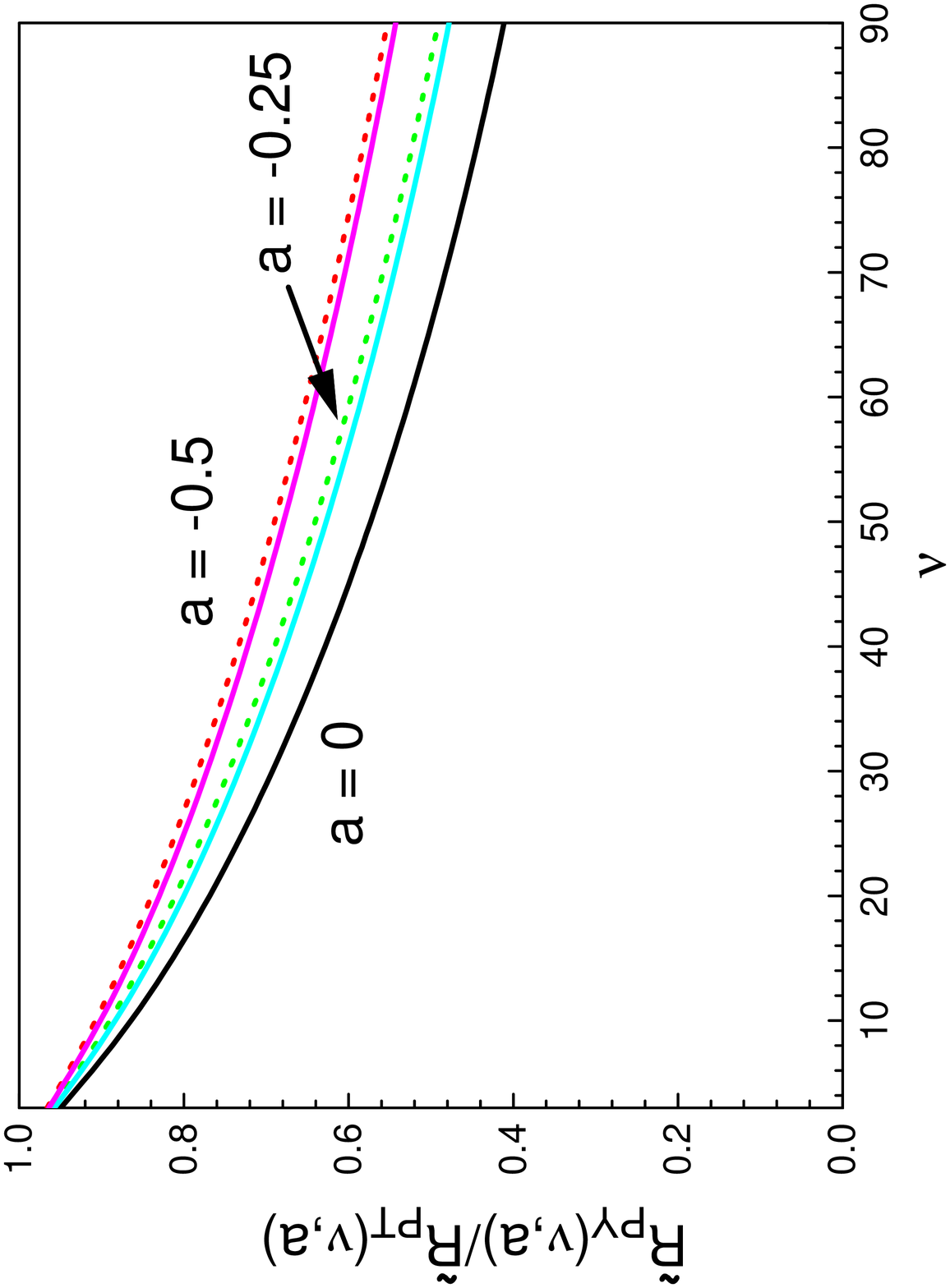,height=11cm,angle=270,clip=0}
{The ratios, Eq.\ (\ref{ratioRs}), between the output of PYTHIA and the NLL resummed
predictions at $a = 0, -1/4, -1/2$ at c.m.\ energy $Q = 91$ GeV.  The solid lines are the directly
computed ratios, the dotted lines is the scaled $a = 0$  curve, according to Eq.\ (\ref{rescale}).\label{Q91} \newline }

\EPSFIGURE{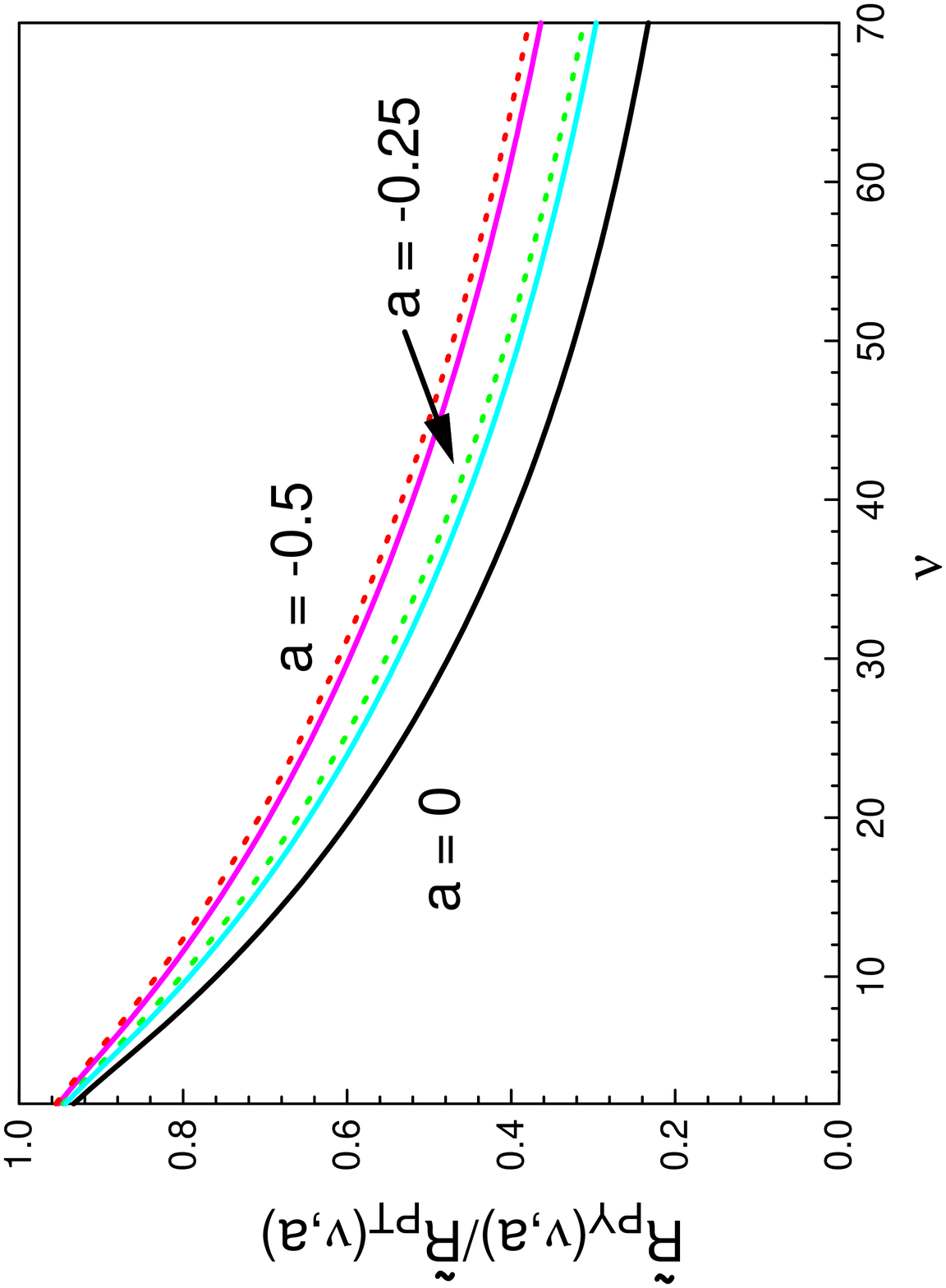,height=11cm,angle=270,clip=0}
{The ratios, Eq.\ (\ref{ratioRs}), between the output of PYTHIA and the NLL resummed
predictions at $a = 0, -1/4, -1/2$ at c.m.\ energy $Q = 35$ GeV, solid and dotted lines as in Fig.
\ref{Q91}.  \label{Q35} \newline}

We note that the logarithms of the curves in both figures 
depend fairly linearly on $\nu$ for relatively small $\nu$, 
indicating that in this range the linear, $\nu/Q$
term dominates.  This is consistent with the result above
for the shifts, which is based only on the $\nu/Q$ correction.
To fit the curves at  larger $\nu$, however, higher powers
in $\nu/Q$ are necessary.  While we have not attempted a fit
of the $a=0$ case (thrust) at different energies our reasoning is
consistent with any determination of the coefficients $\lambda_n(\kappa)$
in Eq.\ (\ref{fdef}) \cite{Korchemsky:2000kp,GarRat1,GarRat2}.

We emphasize that the agreement of our scaling rule with
PYTHIA may only mean  that the output of 
PYTHIA shares some of the properties
that go into the derivation of the rule.

\subsection{Scaling violations}

In the light of the relation between the scaling
rule and boost invariance, Sec.\ \ref{origin},
we can understand why PYTHIA respects the rule
 over a moderate range of  parameter $a$.  
It is natural to think of jet fragmentation
as dynamically boost-invariant along the jet axis, while boost invariance
need not be respected for coherent interjet radiation,
which depends on the relative directions of the jets.  
For moderate values of $a$, the shape functions may follow
the scaling in our numerical tests because in PYTHIA 
the corresponding event weights are
dominated by particles created from boost-invariant dynamics, and for which
correlations between the jet hemispheres (for example, due to decays
into opposite hemispheres) are negligible.  

Beyond NLL, however,
the general resummation for the two-jet event shapes of Eq.\ (\ref{barfdef})
involves coherent interjet radiation \cite{BKuS},
which we expect to produce correlations between
the jet hemispheres.
The neglect of such correlations is
related to the ``inclusive"
approximation discussed in Ref.\ \cite{GarRat1},
in which the effect on the weight function
from off-shell gluons that split into particles
that move into different hemispheres is suppressed.   

More generally, suppose that the correlation
between energy flow is enhanced relative to NLL perturbation theory
for ``short-range" rapidity intervals, 
less than some constant $\Delta \eta_0$.
Then, because of correlations
between radiation in different hemispheres, 
the scaling implied by Eq.\ (\ref{boost})
can hold only when the rapidity range 
for each term on the right-hand 
side is much larger than the constant,
$[n(1-a)]^{-1}\gg \Delta\eta_0$.  For any $\Delta\eta_0$,
this condition is violated for $n$ large enough.
Thus, as $(1-a)$ increases, the scaling is violated, first by 
high powers of $\nu/Q$, and eventually by
lower powers.  Only to the extent that
enhanced short-range correlations are fully negligible can
we expect that the scaling holds for the full
shape function.  Pushing our analysis to larger
values of $(1-a)$ should  eventually
uncover correlations between hemispheres, even in PYTHIA,
perhaps associated with the string breaking picture of
hadronization.  In general, we would expect physical
moments to decrease less rapidly than $(1-a)^{-1}$ in
the presence of positive correlations between
radiation in different hemispheres \cite{BKoS}.

The extension of our analysis 
to multijet events \cite{BSZ} should be relatively straightforward.
In this case, as in the two-jet limit of $\rm e^+e^-$
annihilation, it is necessary to determine
each jet axis by a thrust-like condition, which is
relatively insensitive to recoil effects \cite{Dokshitzer:1998kz,BKuS},
to insure resummation at NLL.  
Such cross sections still factorize perturbatively, but now
into a function involving coherent interjet radiation
as well as jet functions, even at NLL.
The jet functions, but not the soft function,
should obey a scaling relation like Eq.\ (\ref{freln}).
Tests of this scaling for multi-jet shape functions could be an indirect way to
estimate the significance of coherence effects for interjet radiation.  Extensions
to deep-inelastic scattering \cite{dsdis} and hadronic scattering 
\cite{BSZ} may  also be possible.

\section{Conclusions}

We have derived a scaling property
that relates nonperturbative shape functions
within the class of event shapes, including
the thrust, introduced in \cite{BKuS}.
We have seen that the resulting predictions
match the  output of the event generator PYTHIA
over a range of the relevant parameter that
defines the event shape.  This analysis is based on the NLL resummed cross section for
these event shapes, which neglects correlations between hemispheres.
The comparison of these predictions
to actual data should shed light on
universality properties of nonperturbative
corrections, and their relationship to
determinations of $\alpha_s$.
 We hope that this comparison is
still possible for archived data from LEP.

We have argued that the scaling rule derived from the NLL cross section
is more generally dependent on the boost invariance of strong coupling
dynamics.  In general, the scaling must also fail at some level due 
to correlations between hemispheres, associated for example with decays
\cite{BKoS,nasonseymour}.
We might expect such effects to become more important
for $a\ll -1$, since the corresponding weight functions
are sensitive primarily to radiation near the boundary
between the hemispheres.   In any case, we hope that
the example studied above shows 
that event shapes can be designed to 
probe specific aspects of nonperturbative QCD dynamics.

\acknowledgments

We thank Tibor K\'ucs for many useful conversations.
We would like to thank Mike Seymour for a helpful communication. This
work was supported in part by the National Science Foundation grant
PHY-0098527.

\begin{appendix}

\section{Explicit Expressions for the Cross Section in Transform
Space} \label{app:explicit}

For the evaluation of the integrals in Eq.\ (\ref{thrustcomp}) we use
the running coupling at renormalization scale $\mu$ in terms of the 
coupling $\alpha$ evaluated at
  $Q/2$, expanded for use at NLL accuracy,
\ba
\alpha & \equiv & \as\left(\frac{Q}{2} \right)  \label{scaledef} \\
\hspace*{-5mm} \as(\mu) & = & \frac{\alpha}{1 + \frac{\beta_0}{2 \pi}
\alpha \ln \frac{2 \mu}{Q}} \left[ 1 - \frac{\beta_1}{4 \pi \beta_0}
\frac{\alpha}{1 + \frac{\beta_0}{2 \pi} \alpha \ln \frac{2 \mu}{Q}}
\ln \left(1 + \frac{\beta_0}{2 \pi} \alpha \ln \frac{2 \mu}{Q}
\right) + \dots \right], \nonumber \\
& &
\ea
where the coefficients $\beta_0$ and $\beta_1$ are given by
\begin{eqnarray}
\beta_0 & = & \frac{11}{3} C_A - \frac{4}{3} T_F N_f, \label{beta0} \\
\beta_1 & = & \frac{34}{3} C_A^2 - \frac{20}{3} C_A T_F N_f - 4 C_F
T_F N_f. \label{beta1}
\end{eqnarray}
The term with $\beta_1$ is only necessary for the integral containing
$A^{(1)}$ at NLL.

Inserting the expansion of (\ref{scaledef}) into Eq.\ (\ref{thrustcomp}),
the integrals are done to NLL accuracy in terms of elementary
functions by using the replacement 
$\left(\exp[-y]-1\right) \rightarrow-\theta\left(y-{\rm e}^{-\gamma_E}\right)$,
which is accurate to NLL.  Here $\gamma_E$ is the Euler constant.
The result of this procedure is,
\ba
\hspace*{-5mm} \stot  \tilde{\sigma} \left(\nu,Q,a \right) & \equiv &
\left[\,{\mathcal{J}}(\nu,Q,a)\right]^2\nonumber \\
& = & \exp\Bigg\{2 \, \ln (\nu)\, g_1 \left( \frac{\beta_0}{2 \pi}
\frac{\as(\mu)}{2-a} \ln \nu,a\right) \nonumber \\
& & \qquad + \,2 \, \left(\frac{\beta_0}{2 \pi} \right)^2
\frac{\as^2(\mu)}{2-a} \ln^2 \nu\, \ln \left( \frac{2 \mu}{Q} \right)
g_1' \left( \frac{\beta_0}{2 \pi} \frac{\as(\mu)}{2-a} \ln
\nu,a\right) \nonumber \\
& & \qquad + \,2 \,g_2 \left( \frac{\beta_0}{2 \pi}
\frac{\as(\mu)}{2-a} \ln \nu,a\right) + {\mathcal{O}} \left( \as^n
\ln^{n-1} \nu \right) \Bigg\}, \label{expgrun}
\ea
where the functions $g_1$ and $g_2$ and $g'_1$ that resum leading and
next-to-leading logarithms, respectively, are given by
\ba
g_1(x,a) & = & - \frac{4}{\beta_0} \frac{1}{1-a} \frac{1}{x} A^{(1)}
\Bigg[ \left( \frac{1}{2-a} - x \right) \ln (1-(2-a) x)  - \, (1-x)
\ln (1-x) \Bigg]  \nonumber  \\
& & \label{g1} \\
g_2(x,a) & = & \frac{2}{\beta_0} B^{(1)} \ln (1-x)  -
\frac{8}{\beta_0^2} \frac{1}{1-a}  A^{(2)} \left[ (2-a) \ln (1-x) -
\ln (1-(2-a)x) \right] \nonumber \\
& &  - \, \frac{4}{\beta_0} \gamma_E \frac{1}{1-a}  A^{(1)} \left[
\ln (1-x) - \ln(1-(2-a) x)\right] \nonumber \\
& &  + \, \frac{4}{\beta_0} \ln 2 \frac{1}{1-a} \, A^{(1)} \left[
(2-a) \ln (1-x) -  \ln(1-(2-a) x)\right] \nonumber \\
& &  - \frac{\beta_1}{\beta_0^3} \frac{1}{1-a} A^{(1)} \left[ 2
\ln(1-(2-a) x) - 2(2-a) \ln(1-x) \right. \nonumber \\
& & \qquad \qquad \qquad + \left. \, \ln^2 (1-(2-a) x) - (2-a) \ln^2
(1-x) \right] \label{g2}  \\
g_1' \left( x,a\right) & = & \frac{\partial }{\partial x} g_1 (x,a) .
\label{g1prime}
\ea
  In Eq.\ (\ref{expgrun}) the scale $\mu$ should be chosen
of the order of the hard scale to avoid further large logarithms, and
   we choose $\mu=Q$.
Setting $\mu = Q$ as in \cite{CTTW2} cancels the term proportional to
$\ln 2$ in Eq.\ (\ref{g2}), and we reproduce for $a = 0$ the form of
\cite{CTTW2}.

Finally, for sake of completeness,
we list the coefficients $G_{ij}$ that occur in the expansion
in terms of $\as$ of the logarithm of the resummed radiator, Eq. (\ref{resumexp}),
\ba
G_{11}(a) & = & \frac{3}{2-a} C_F, \\
G_{12}(a) & = & -\frac{2}{2-a} C_F, \\
G_{22}(a) & = & - \frac{1}{(2-a)^2} \frac{1}{36} C_F \left[ 48 \pi^2 C_F + \left(169-134 \,a - 6\,(2-a) \pi^2\right)
C_A \right. \nonumber \\
& & \qquad \qquad \qquad \qquad \left. - 2\, (11-10 \,a) N_f \right], \\
G_{23}(a) & = & - \frac{3-a}{(2-a)^2} \frac{1}{9} C_F \left(11 C_A - 2 N_f \right).
\ea
For $a = 0$, the $G_{ij}$s again reduce to those listed in \cite{CTTW2}.

\end{appendix}


\begin{thebibliography}{00}


\bibitem{Farhi:1977sg}
E.~Farhi, Phys.\ Rev.\ Lett.\  {\bf 39}, 1587 (1977).

\bibitem{weights} 
H.~Georgi and M.~Machacek,
Phys.\ Rev.\ Lett.\  {\bf 39}, 1237 (1977); \\
G.~Parisi,
Phys.\ Lett.\ B {\bf 74}, 65 (1978); \\
J.~F.~Donoghue, F.~E.~Low and S.~Y.~Pi,
Phys.\ Rev.\ D {\bf 20}, 2759 (1979); \\
G.~C.~Fox and S.~Wolfram,
Nucl.\ Phys.\ B {\bf 149}, 413 (1979)
[Erratum-ibid.\ B {\bf 157}, 543 (1979)].


\bibitem{broad1} 
S.~Catani, G.~Turnock and B.~R.~Webber, 
Phys.\ Lett.\ B {\bf 295}, 269 (1992).

\bibitem{heavyjet} 
T.~Chandramohan and L.~Clavelli,
Nucl.\ Phys.\ B {\bf 184}, 365 (1981).

\bibitem{eflow}
C.~L.~Basham, L.~S.~Brown, S.~D.~Ellis and S.~T.~Love,
Phys.\ Rev.\ D {\bf 17}, 2298 (1978); \\
N.~A.~Sveshnikov and F.~V.~Tkachov,
Phys.\ Lett.\ B {\bf 382}, 403 (1996)
[\hepph{9512370}]; \\
F.~V.~Tkachov, 
Int.\ J.\ Mod.\ Phys.\ A {\bf 12}, 5411 (1997)
[\hepph{9601308}]; \\
G.~P.~Korchemsky, G.~Oderda and G.~Sterman, in {\it 5th International 
Workshop on Deep Inelastic Scattering and QCD (DIS 97)}, AIP 
Conference Proceedings 407,
ed.\ J.\ Repond, D.\ Krakauer (American Institute of Physics, 
Woodbury, NY 1978), p.\ 988
[\hepph{9708346}]; \\
C.~F.~Berger {\it et al.},
in {\it Proc. of the APS/DPF/DPB Summer Study on 
the Future of Particle Physics (Snowmass 2001) } ed. N.~Graf,
eConf {\bf C010630}, P512 (2001)
[\hepph{0202207}].

\bibitem{irren}
H.~Contopanagos and G.~Sterman, Nucl.\ Phys.\ B {\bf 419}, 77 (1994) 
[\hepph{9310313}]; \\
B.~R.~Webber, Phys.\ Lett.\ B {\bf 339}, 148 (1994)
[\hepph{9408222}]; \\
Y.~L.~Dokshitzer and B.~R.~Webber, Phys.\ Lett.\ B {\bf 352}, 451 (1995)
[\hepph{9504219}]; \\
R.~Akhoury and V.~I.~Zakharov, Nucl.\ Phys.\ B {\bf 465}, 295 (1996)
[\hepph{9507253}]; \\
M.~Beneke, V.~M.~Braun and L.~Magnea, Nucl.\ Phys.\ B {\bf 497}, 297 (1997)
[\hepph{9701309}]; \\
M. Beneke and V.M.\ Braun, in the Boris Ioffe Festschrift, \textit{At 
the Frontier of
Particle Physics / Handbook of QCD}, ed.\ M. Shifman (World 
Scientific, Singapore, 2001), vol.
3, p.\ 1719 [\hepph{0010208}].


\bibitem{Manohar:1994kq}
A.~V.~Manohar and M.~B.~Wise, Phys.\ Lett.\ B {\bf 344}, 407 (1995)
[\hepph{9406392}].

\bibitem{irrdiff} G.~P.~Korchemsky and G.~Sterman, Nucl.\ Phys.\ B 
{\bf 437}, 415 (1995)
[\hepph{9411211}]; \\
G.~P.~Korchemsky and G.~Sterman, in \textit{Moriond 1995}: Hadronic:0383-392
[\hepph{9505391}].


\bibitem{DokWeb}
Y.~L.~Dokshitzer and B.~R.~Webber,
Phys.\ Lett.\ B {\bf 404}, 321 (1997)
[\hepph{9704298}].

\bibitem{dispers} Y.~L.~Dokshitzer, G.~Marchesini and B.~R.~Webber,
Nucl.\ Phys.\ B {\bf 469}, 93 (1996)
[\hepph{9512336}]; \\
Y.~L.~Dokshitzer, A.~Lucenti, G.~Marchesini and G.~P.~Salam,
Nucl.\ Phys.\ B {\bf 511}, 396 (1998)
[Erratum-ibid.\ B {\bf 593}, 729 (2001)]
[\hepph{9707532}];\\
Y.~L.~Dokshitzer, A.~Lucenti, G.~Marchesini and G.~P.~Salam,
JHEP {\bf 9805}, 003 (1998)
[\hepph{9802381}];\\
Y.~L.~Dokshitzer, G.~Marchesini and G.~P.~Salam,
Eur.\ Phys.\ J.\ direct C {\bf 1}, 3 (1999)
[\hepph{9812487}];\\
Y.~L.~Dokshitzer, G.~Marchesini and B.~R.~Webber,
JHEP {\bf 9907}, 012 (1999)
[\hepph{9905339}];\\
M.~Dasgupta, L.~Magnea and G.~Smye,
JHEP {\bf 9911}, 025 (1999)
[\hepph{9911316}].


\bibitem{oneqdisrefs}  
A.~H.~Mueller,
Phys.\ Lett.\ B {\bf 308}, 355 (1993);
\\
M.~Dasgupta and B.~R.~Webber,
Phys.\ Lett.\ B {\bf 382}, 273 (1996)
[\hepph{9604388}]; \\
M.~Dasgupta and B.~R.~Webber, 
Eur.\ Phys.\ J.\ C {\bf 1}, 539 (1998)
[\hepph{9704297}]; \\
M.~Dasgupta and B.~R.~Webber, 
JHEP {\bf 9810}, 001 (1998)
[\hepph{9809247}].

\bibitem{oneqdish1} 
C.~Adloff {\it et al.}  [H1 Collaboration],
Phys.\ Lett.\ B {\bf 406}, 256 (1997)
[\hepex{9706002}].


\bibitem{alphasmean} 
S.~Bethke,
J.\ Phys.\ G {\bf 26}, R27 (2000)
[\hepex{0004021}]; \\
S.~J.~Burby and C.~J.~Maxwell, 
Nucl.\ Phys.\ B {\bf 609}, 193 (2001)
[\hepph{0011203}]; \\
G.~Dissertori, \textit{Measurements of alpha(s) from event shapes and the four-jet rate},
[\hepex{0209070}].

\bibitem{MovillaFernandez:2001ed}
P.~A.~Movilla Fernandez, S.~Bethke, O.~Biebel and S.~Kluth, Eur.\
Phys.\ J.\ C {\bf 22}, 1 (2001) [\hepex{0105059}].


\bibitem{CTTW1}
S.~Catani, G.~Turnock, B.~R.~Webber and L.~Trentadue, Phys.\ Lett.\ B
{\bf 263}, 491 (1991).

\bibitem{CTTW2}
S.~Catani, L.~Trentadue, G.~Turnock and B.~R.~Webber, Nucl.\ Phys.\ B
{\bf 407}, 3 (1993).

\bibitem{Dokshitzer:1998kz}
Y.~L.~Dokshitzer, A.~Lucenti, G.~Marchesini and G.~P.~Salam, JHEP
{\bf 9801}, 011 (1998) [\hepph{9801324}].

\bibitem{BKuS}
C.~F.~Berger, T.~K\'ucs and G.~Sterman, \textit{Interjet energy flow
/ event shape correlations}, \hepph{0212343}; \\
C.~F.~Berger, T.~K\'ucs and G.~Sterman,
Phys.\ Rev.\ D {\bf 68}, 014012 (2003)
[\hepph{0303051}].

\bibitem{dsdis} 
M.~Dasgupta and G.~P.~Salam, 
JHEP {\bf 0208}, 032 (2002)
[\hepph{0208073}]; \\
M.~Dasgupta and G.~P.~Salam, 
Eur.\ Phys.\ J.\ C {\bf 24}, 213 (2002)
[\hepph{0110213}].

\bibitem{dsnonglo}  M.~Dasgupta and G.~P.~Salam,
Phys.\ Lett.\ B {\bf 512}, 323 (2001)
[\hepph{0104277}].


\bibitem{KorSt99}
G.~P.~Korchemsky and G.~Sterman, Nucl.\ Phys.\ B {\bf 555}, 335
(1999) [\hepph{9902341}].


\bibitem{KorMor98}
G.~P.~Korchemsky, \textit{Shape functions and power corrections to
the event shapes}, in \textit{Minneapolis 1998, Continuous advances
in QCD}, 179 (1998) [\hepph{9806537}].

\bibitem{Korchemsky:2000kp}
G.~P.~Korchemsky and S.~Tafat, JHEP {\bf 0010}, 010 (2000)
[\hepph{0007005}].

\bibitem{BKoS} A.~V.~Belitsky, G.~P.~Korchemsky and G.~Sterman,
Phys.\ Lett.\ B {\bf 515}, 297 (2001) [\hepph{0106308}].

\bibitem{GarRat1} E.~Gardi and J.~Rathsman, Nucl.\ Phys.\ B {\bf 
609}, 123 (2001)
[\hepph{0103217}].

\bibitem{GarRat2} E.~Gardi and J.~Rathsman, Nucl.\ Phys.\ B {\bf 
638}, 243 (2002)
[\hepph{0201019}].

\bibitem{GarMag} E.~Gardi and L.~Magnea, \textit{The C parameter 
distribution in $e^+ e^-$ annihilation},
\hepph{0306094}.

\bibitem{BSZ} 
A.~Banfi, Y.~L.~Dokshitzer, G.~Marchesini and G.~Zanderighi,
JHEP {\bf 0103}, 007 (2001)
[\hepph{0101205}]; \\
A.\ Banfi, G.P.\ Salam and G.\
Zanderighi, \textit{Generalized resummation of QCD final-state observables},
\hepph{0304148}.

\bibitem{Sjostrand:2000wi}
T.~Sjostrand, P.~Eden, C.~Friberg, L.~Lonnblad, G.~Miu, S.~Mrenna and
E.~Norrbin, Comput.\ Phys.\ Commun.\  {\bf 135}, 238 (2001)
[\hepph{0010017}].

\bibitem{Salam:2001bd}
G.~P.~Salam and D.~Wicke, JHEP {\bf 0105}, 061 (2001) [\hepph{0102343}].

\bibitem{Sterman:2003wk}
G.~Sterman, \textit{Approaching the final state in perturbative QCD},
\hepph{0301243}.

\bibitem{Catani:1996jh}
S.~Catani and M.~H.~Seymour, Phys.\ Lett.\ B {\bf 378}, 287 (1996)
[\hepph{9602277}].

\bibitem{eikonalrules} 
N.~Kidonakis, G.~Oderda and G.~Sterman,
Nucl.\ Phys.\ B {\bf 531}, 365 (1998)
[\hepph{9803241}].

\bibitem{eikexp} G.~Sterman, 
in \textit{AIP Conference Proceedings Tallahassee, Perturbative Quantum Chromodynamics}, 
eds. D. W. Duke, J. F. Owens, New York, 1981; \\
J.~G.~Gatheral, Phys.\ Lett.\ B {\bf 133}, 90 (1983); \\
J.~Frenkel and J.~C.~Taylor, Nucl.\ Phys.\ B {\bf 246}, 231 (1984); \\
G.~P.~Korchemsky and A.~V.~Radyushkin,
Phys.\ Lett.\ B {\bf 171}, 459 (1986);\\
C.~F.~Berger, Phys.\ Rev.\ D {\bf 66}, 116002 (2002) [\hepph{0209107}].

\bibitem{Berger:2003zh}
C.~F.~Berger, \textit{Soft gluon exponentiation and resummation},
Ph.D. Thesis, SUNY at Stony Brook, May 2003, \hepph{0305076}.

\bibitem{nasonseymour}
P.~Nason and M.~H.~Seymour, Nucl.\ Phys.\ B {\bf 454}, 291 (1995)
[\hepph{9506317}].


\bibitem{Sjostrand:2001yu}
T.~Sjostrand, L.~Lonnblad and S.~Mrenna, \textit{PYTHIA 6.2:
Physics and manual}, [\hepph{0108264}].

\bibitem{Andersson}
B.~Andersson, G.~Gustafson, G.~Ingelman and T.~Sjostrand, Phys.\
Rept.\  {\bf 97}, 31 (1983); \\
B.~Andersson, \textit{The Lund Model}, Cambridge Monogr.\ Part.\ Phys.\
Nucl.\ Phys.\ Cosmol.\  {\bf 7}, 1 (1997).



\end{thebibliography}
\end{document}